\documentclass[epj,twocolumn]{svjour}

\RequirePackage[T1]{fontenc}

\RequirePackage{graphicx}
\RequirePackage{mathptmx}      
\RequirePackage{flushend}
\RequirePackage[numbers,sort&compress]{natbib}
\RequirePackage[colorlinks,citecolor=blue,urlcolor=blue,linkcolor=blue]{hyperref}

\journalname{Eur. Phys. J. C}

\usepackage{graphicx}
\usepackage{dcolumn}
\usepackage{bm}
\usepackage{booktabs}

\usepackage{bm}
\usepackage{longtable}
\usepackage[usenames,dvipsnames]{xcolor}
\usepackage{multirow}
\usepackage{graphicx}
\usepackage{hyperref}
\usepackage{times}
\usepackage{color}
\usepackage{breakurl}
\usepackage{mathrsfs}

\usepackage{appendix}

\def\be{\begin{equation}}
\def\ee{\end{equation}}
\def\bea{\begin{eqnarray}}
\def\eea{\end{eqnarray}}
\def\mnras{Mon. Not. Roy. Astr. Soc.}
\def\apj{Astrophys. J.}

\def\nphysa{Nucl. Phys. A}
\def\jcap{JCAP}
\def\apjs{ApJ Suppl.}
\def\physrep{Phys. Rep.}

\begin{document}

\title{Neutrino oscillation in the $q$-metric}

\author{Kuantay Boshkayev${}^{1,2,3}$\thanks{email: kuantay@mail.ru, kuantay.boshkayev@nu.edu.kz} \and Orlando Luongo${}^{1,4}$\thanks{email: orlando.luongo@lnf.infn.it} \and Marco Muccino${}^{1,5}$\thanks{email: marco.muccino@lnf.infn.it}}

\institute{National Nanotechnology Laboratory of Open Type, Department of Theoretical and Nuclear Physics, Al-Farabi Kazakh National University, 
050040 Almaty, Kazakhstan. \and Department of Physics, Nazarbayev University, 
010000 Nur-Sultan, Kazakhstan. \and Fesenkov Astrophysical Institute, 
050020 Almaty, Kazakhstan \and Scuola di Scienze e Tecnologie, Universit\`a di Camerino, 62032 Camerino, Italy. \and Istituto Nazionale di Fisica Nucleare (INFN), Laboratori Nazionali di Frascati, 00044 Frascati, Italy.}

\abstract{
We investigate neutrino oscillation in the field of an axially symmetric space-time, employing the so-called $q$-metric, in the context of general relativity. Following the standard approach, we compute the phase shift  invoking the weak and strong field limits and small deformation. To do so, we consider neutron stars, white dwarfs and supernovae as strong gravitational regimes whereas the Solar System as weak field regime. We argue that the inclusion of  the quadrupole parameter leads to the modification of the well-known results coming from the spherical solution due to the  Schwarschild space-time. Hence, we show that in the Solar System regime, considering the Earth and Sun, there is a weak probability to detect deviations from the flat case, differently from the case of neutron stars and white dwarfs in which this probability is larger. Thus, we heuristically discuss some implications on constraining the free parameters of the phase shift by means of  astrophysical neutrinos. A few consequences in cosmology and possible applications for future space experiments are also discussed throughout the text.
}


\maketitle

\section{Introduction}\label{sezione1}

Ever since their discovery \cite{preliminare1,preliminare2}, neutrinos have been under scrutiny for their exotic and enigmatic properties. In the standard model of particle physics, neutrinos are massless and left-handed particles, albeit recent observations definitively showed that these particles have a non-vanishing mass \cite{due,uno2,uno1}.

On the one hand, the absolute scale of neutrino's mass spectra is yet unknown, although on the other hand the minimum scale\footnote{The mixing angles associated with atmospheric and Solar transitions are so far quite large leading to unbounded results.} is given by the larger mass splitting, set around $\sim50$ meV \cite{tre}. Both flavor mixing and neutrino oscillation are therefore theoretical challenges for quantum field theory since Pontecorvo's original treatment in which the phenomenon of oscillation was firstly described\footnote{The original proposal for massive neutrino mixing and oscillation has been argued in flat space-time \cite{quattrotris}. Indirect evidence for massive neutrinos comes from the Solar neutrino deficit, the atmospheric neutrino anomaly and the evidence from the LSND experiment \cite{quattrobis}. } \cite{quattro}.

Immediately after having introduced the concept of neutrino oscillation, Mikheyev, Smirnov and Wolfenstein investigated transformations of one neutrino flavor into another in media with  non-constant density \cite{cinquebis,cinque}. To understand the origin of neutrino masses, possible extensions of the standard model of particle physics have been extensively reviewed~\cite{sei}, whereas several experimental studies have fixed bounds on atmospheric and Solar neutrino oscillation. The theoretical scheme behind oscillation has been widely investigated so far, giving rise to a wide number of different treatments and approaches \cite{oscy} to disclose  the origin of neutrino masses.

In particular, an intriguing challenge is to understand the role played by strong gravitational fields on oscillation \cite{stodolsky}. In fact, when the effects of gravity are not negligible, one is forced to use curved space-times in general relativity (GR) to characterize how matter's distribution influences the oscillation itself \cite{curvo2bis,curvo2}. In this respect, neutrino oscillation in curved space-times has been reviewed under several prescriptions and conceptually one can consider two main perspectives intimately interconnected between them. The first interpretation assumes curved geometry \emph{to fuel the oscillation}. Here, space-time behaves as \emph{a source} for neutrino oscillation, whereas the second one assumes that oscillation \emph{is modified by gravity}, without being a pure source (see e.g. \cite{Ioscy,Ioscybis,IIoscy,refe1}). The former does not act as a source for the oscillation itself. Obviously, the two approaches do not show a strong dichotomy since the gravitational contribution is responsible for oscillation in both the cases.

Both interpretations, although appealing, are so far theoretical speculations only in which the neutrino phase shift can be computed once the space-time symmetry is assumed \emph{a priori} \cite{sol1,sol1tris,sol1bis,sol1quinquis,wudka,sol2bis,sol1quatris,sol2tris,sol2quatris,sol2quinquis,sol3,sol3bis}. In particular, there exists a number of exact and approximate solutions of Einstein's field equations \cite{solutions, bqr12} capable of matching the neutrino oscillations with the space-time symmetry \cite{sol4,sol5}. Recent developments have prompted that the effects of rotation for spherically symmetric space-time can be neglected in the weak field and slow rotation regimes \cite{sol4}, especially for the Solar and atmospheric neutrinos. The effects of deformation, then, could be of interest even for Earth and stars but also for astronomical compact objects, such as neutron stars (NSs) and white dwarfs (WDs). Even though different metrics can be used to describe such configurations, we here focus on the simplest space-time departing from a pure spherical symmetry by adding a deformation term, i.e. the Zipoy-Voorhees space-time. For the sake of completeness, the  investigation of the effects of rotation the Hartle-Thorne metric is also involved   \cite{1967ApJ...150.1005H,1968ApJ...153..807H}.

In this work we therefore take into account the  exterior and well-consolidated Zipoy-Voorhees metric, often termed in the literature as gamma-metric, delta-metric or more frequently $q$-metric\footnote{It describes a  static and deformed astrophysical object whose gravitational field generalizes the Schwarzschild metric through the inclusion of a quadrupole term \cite{quev12}.} \cite{mala04,mala05,Mashhoon2018,quev11}. So, motivated by the fact that the $q$-metric is able to model exteriors of several compact objects, we investigate the corresponding consequences of neutrino oscillation. To do so, we consider both weak and strong field regimes with small deformation of the source. Thus, we evaluate the phase shift and define the additional terms that modify the shift with respect to the case of Schwarzschild space-time. Afterwards, we apply our results to astrophysical cases, i.e. to those compact objects which exhibit a spherical symmetry. In this respect, we involve WDs and NSs and we justify why supernovae, and well-consolidate standard candles in general, are unable for being indicators of neutrino production, through the fact that neutrinos are produced from the NSs born out of the explosion and the corresponding  oscillation is also affected by predominant matter effects. In particular, by means of experimental data from cosmological probes and nuclear physics experiments,  within the current paradigm purporting the three-flavor neutrino mixing theory, we compute numerical constraints on survival probability for WDs and NSs, respectively weak and strong gravitational fields. We compare our results with previous expectations, concluding  that on the Earth the quadrupole moment effect is negligible, albeit in the case of rotating WDs and NSs it affects the mass difference $\Delta m_{21}^2$ between neutrino eigenstates 1 and 2. We analyze the dipole and quadrupole cases in view of maximally-rotating configurations and we compare our findings with the ones computed in the Hartle-Thorne (HT) space-time. We give hints toward possible experiments to be performed in the next years to check the theoretical deviations here developed. Finally we produce a set of numerical bounds which agree and extend the outcomes of previous works.

The paper is structured as follows. In Sec. \ref{sezione2} we give details on the axially symmetric $q$-metric and on its principal properties. There, we include details on motion of test particles explicitly reporting the space-time kinematics. In Sec. \ref{phaseshift}, we give a fully-detailed explanation on neutrino phase shift first and then we specialize it to the case of the  $q$-metric. The neutrino oscillation is thus faced and the monopole, dipole and quadrupole corrections are explicitly reported and commented, particularly for the HT metric. Afterwards, we pass through Sec. \ref{excons} in which we give a brief summary of the current status-of-art of numerical constraints over neutrino masses. We consider separately the cosmological case from Solar system constraints and reactor bounds. Then, we give our computational bounds which have been summarized in the corresponding tables. In the same part, we emphasize the physical reasons for adopting compact objects, such as NSs and WDs as benchmarks for investigating neutrino oscillation in space. The case of weakly and strongly interacting gravitational fields are summarized respectively in Sec. \ref{wdns}, in which we analyze NSs as strong gravitational landscape and WDs as weak gravitational scenario for neutrino oscillation. Finally, in Sec. \ref{gedanken}, we discuss the theoretical consequences of our approach and the corresponding experimental developments, emphasizing a possible \emph{Gedankenexperiment} in which future experiments can be calibrated. The last part of the work, Sec. \ref{concl} presents final outlooks and perspectives.


\section{Quasi axisymmetric space-time}
\label{sezione2}

We here start handling  axisymmetric space-times highlighting their general properties. To do so, let us first consider the Weyl class of static axisymmetric vacuum solutions. In particular, by means of prolate spheroidal coordinates, namely $(t,x,y,\phi)$, with $x\geq 1$  and $-1\leq y \leq1$, the class of solutions is defined by
\begin{equation}\label{eq:metgen}
ds^2 =-e^{2\psi} dt ^2 +\bar{m}^2\Big[ f_1\left( \frac{dx^2}{x^2-1} + \frac{dy^2}{1-y^2} \right) + f_2d\phi^2\Big],
\end{equation}
where
\begin{eqnarray}
\bar{m}^2&\equiv& m^2 e^{-2\psi}\,,\\
f_1&\equiv& e^{2\gamma}(x^2-y^2)\,,\\
f_2&\equiv& (x^2-1)(1-y^2)\,,
\end{eqnarray}
with $\psi=\psi(x,y)$ and $\gamma=\gamma(x,y)$, functions of spatial coordinates only while $m$ represents the standard mass parameter.

The correspondence between such a class of models, Eq. (\ref{eq:metgen}), and spherical coordinates for the well-known Schwarzschild solution is found as
\be
\label{schw}
\psi_S=\frac12\ln{\frac{x-1}{x+1}}\ , \qquad
\gamma_S=\frac12\ln{\frac{x^2-1}{x^2-y^2}}\,.
\ee
Here, the functions $\psi$ and $\gamma$ may be generalized by means of the Zipoy \cite{zip66} and Voorhees \cite{voor70} transformation once a seed (Schwarzschild) solution is known. Hence, the Zipoy-Voorhees generalization of the Schwarzschild solution
\be\label{eqZP}
\psi = \frac{\delta}{2}\ln\frac{x-1}{x+1}\ , \quad
\quad \gamma = \frac{\delta^2}{2}\ln \frac{x^2-1}{x^2-y^2}\,,
\ee
in which the parameter $\delta$ can be written by
\begin{equation}
    \delta=1+q\,,
\end{equation}
where $q$ represents the deformation parameter of the source, or alternatively, the quadrupole parameter. Hence the Zipoy-Voorhees metric is often referred to as the \emph{$q$-metric} to stress the role played by $q$. In doing so, the $q$-metric is defined by the line element Eqs. (\ref{eq:metgen}) and (\ref{eqZP}), with the requirement that $\delta=1+q$. When $q$ vanishes, the $q$-metric reduces to the Schwarzschild solution. We are interested in employing Eqs. (\ref{eq:metgen}) and (\ref{eqZP}) to compute neutrino oscillation. To perform this, let us first consider the dynamical consequences of Eqs. (\ref{eq:metgen}) and (\ref{eqZP}) in the next subsection.

\subsection{Motion of test particles}

The geodesics of test particles  in the $q$-metric are the key ingredients to calculate the neutrino phase shift. To evaluate the test particle motion we follow the standard procedure \cite{bglq,Herrera2000IJMPD,Chowdhury2012,Boshkayev2016,2019PhRvD..99d4012B,2019arXiv190406207A} and particularly  by considering Killing symmetries and the normalization conditions  $g_{\alpha\beta}\dot x^\alpha\dot x^\beta=-\mu^2$, one can assume $E$ and $L$ as the conserved energy and angular momentum respectively. As usual, $E$ and $L$ are associated with the Killing vectors $\partial_t$ and  $\partial_\phi$ respectively of the test particle. Assuming that $\mu$ is the particle mass, we have
\begin{eqnarray}
\label{geoeqns}
&\dot t=\frac{E}{e^{2\psi}}\ , \qquad
\dot \phi=\frac{L e^{2\psi}}{m^2X^2Y^2}\ ,\nonumber\\
&\ddot y=-\frac{Y^2}{X^2}\left[\zeta_y+\frac{y}{X^2+Y^2}\right]{\dot x}^2+2\left[\zeta_x-\frac{x}{X^2+Y^2}\right]{\dot x}{\dot y}\nonumber\\
&+\left[\zeta_y-\frac{y}{X^2+Y^2}\frac{X^2}{Y^2}\right]{\dot y}^2-e^{-2\gamma}\frac{Y^2[L^2 e^{4\psi}+E^2m^2X^2Y^2]\psi_y+yL^2e^{4\psi}}{m^4X^2Y^2(X^2+Y^2)}\ ,\nonumber\\
&{\dot x}^2=-\frac{X^2}{Y^2}{\dot y}^2+\frac{e^{-2\gamma}X^2}{m^2(X^2+Y^2)}\left[E^2-\mu^2e^{2\psi}-\frac{L^2e^{4\psi}}{m^2X^2Y^2}\right]\,,\nonumber
\end{eqnarray}
where $\zeta\equiv\psi-\gamma$ and the dot indicates the derivative with respect to the affine parameter $\lambda$ along the curve, whereas
\begin{eqnarray}
X=\sqrt{x^2-1}\,,\qquad Y=\sqrt{1-y^2}\,.
\end{eqnarray}

\noindent In particular $\lambda$ is the proper time for time-like geodesics by setting $\mu=1$.
Furthermore, since $U^\alpha U_\alpha=-1$, we can associate $U^\alpha$ with 4-velocity vector, while the additional case of null geodesics are characterized instead by $\mu=0$ and $K^\alpha  K_\alpha=0$, which is now a tangent vector. The simplest approach deals with the motion on the symmetry plane $y=0$. In particular, let us notice that if both $y$ and $\dot y$ vanish, then the $3^{rd}$ equation of Eqs. (\ref{geoeqns}) shows the motion is located inside the symmetry plane only. This happens as all the derivatives  with respect to $y$ of the functions defined in the $q$-metric are zero\footnote{Relaxing the hypothesis $y=0$ would lead to  modifications in geodesics, which we expect to depend upon the value of $q$. The larger $q$, the greater changes are expected, showing always more complicated expressions whose integration will be possible only numerically. This motivates the simplest choice of lying at the symmetry plane.}. \noindent Finally, considering the relation between our space-time and  Boyer-Lindquist coordinates
$(t,r,\theta,\phi)$ is given by $t=t, x=\frac{r-M}{\sigma}, y=\cos\theta, \phi=\phi$, Eqs. (\ref{geoeqns}) simply read:
\bea
\label{geoeqnsequat}
\dot t&=&\frac{E}{e^{2\psi}}\nonumber\\
\dot \phi&=&\frac{Le^{2\psi}}{\sigma^2X^2}\ , \\
{\dot x}&=&\pm\frac{e^{-\gamma}X}{m\sqrt{1+X^2}}\left[E^2-\mu^2e^{2\psi}-\frac{L^2e^{4\psi}}{m^2X^2}\right]^{1/2}\ ,\nonumber
\eea
in which every parameter has been computed with the prescription $y=0$. In view of these results, we are now ready to compute the phase shift for neutrino oscillation in the next section.

\section{The neutrino phase shift in the $q$-metric}\label{phaseshift}

The phase associated with neutrinos of different mass eigenstate \cite{stodolsky} is given by
\be
\label{phase}
\Phi_k=\int_A^BP_{\mu\,(k)} dx^\mu\,.
\ee
This phase is associated with the 4-momentum $P=m_kU$ of a given neutrino that \emph{is produced} at a precise space-time point, namely $A$, and \emph{is detected} at another defined space-time point, say $B$.

The standard assumptions \cite{petcov} usually applied to the evaluation of the phase are enumerated below.
\begin{itemize}
    \item[{\bf I.}\,\,\,\,\,] A massless trajectory is taken into account.
    \item[{\bf II.}\,\,] The mass eigenstates are energy eigenstates.
    \item[{\bf III.}] The ultrarelativistic approximation is valid, i.e. $m_k\ll E$.
\end{itemize}
The conditions above reported imply respectively that

\begin{itemize}
    \item neutrinos travel along  null geodesic paths;
    \item neutrino eigenstates  have all a common energy, say $E$;
    \item all quantities are evaluated up to first order in $m_k/E$.
\end{itemize}

\noindent Thus, the integral is carried out over a null path, so that Eq. (\ref{phase}) can be also written as
\be
\label{phase2}
\Phi_k=\int_{\lambda_A}^{\lambda_B}P_{\mu\,(k)}K^\mu d\lambda\ ,
\ee
where $K$ is a null vector tangent to the photon path.
The components of $P$ and $K$ are thus obtained from Eq. (\ref{geoeqnsequat}) by setting $\mu=m_k$ and $\mu=0$ respectively.
In the case of equatorial motion the argument of the integral in Eq. (\ref{phase2}) depends on the coordinate $x$ only, so that the integration over the affine parameter $\lambda$ can be switched over $x$ by
\be
\label{phase2equat}
\Phi_k=\int_{x_A}^{x_B}P_{\mu\,(k)}\frac{K^\mu}{K^x} dx\ ,
\ee
where $K^x= dx/d\lambda$.
By applying the relativistic condition $m_k\ll E$ we find
\be
\label{phase3}
\Phi_k\simeq\mp\frac{m_k^2}{2E}\int_{x_A}^{x_B}\frac{\sigma^2 xe^{\gamma}}{\sqrt{\sigma^2(x^2-1)-e^{2\psi}(b-\omega)^2}} dx\ ,
\ee
to first order in the expansion parameter $m_k/E\ll1$, where $E$ is the energy for a massless neutrino and $b=L/E$ the impact parameter. Hence, the phase shift $\Phi_{kj}\equiv\Phi_k-\Phi_j$ responsible for the oscillation is given by
\be
\label{shift}
\Phi_{kj}
\simeq\mp\frac{\Delta m_{kj}^2}{2E}\int_{x_A}^{x_B}\frac{\sigma^2xe^{\gamma}}{\sqrt{\sigma^2(x^2-1)-e^{2\psi}(b-\omega)^2}} dx\ ,
\ee
where $
\Delta m_{kj}^2=m_k^2-m_j^2$.

\subsection{Neutrino oscillations in the $q$-metric}\label{neutosqm}

The exterior field of a deformed object is described by the $q$-metric \cite{quev11,quev12,quev13}, whose line element can be written in the Lewis-Papapetrou form, Eq. (\ref{eq:metgen}), by means of
\be
\psi = (1+q)\psi_S\ , \quad
\quad \gamma = (1+q)^2\gamma_S\ ,
\ee
where $\psi_S$ and $\gamma_S$ are given\footnote{The Schwarzschild solution corresponds to $q\rightarrow0$.} by Eq.~(\ref{schw}).
Then the phase shift, Eq. (\ref{shift}), expressed in terms of the standard spherical coordinates $(t, r, \theta, \phi)$,  for the $q$-metric becomes
\bea
\Phi_{kj}\simeq \mp\frac{\Delta m_{kj}^2}{2E}
\left[r+\frac{m^2}{r}\left\{q-\frac{b^2}{2m^2}+\frac{2q(b^2+m^2)+b^2}{2mr}\right\}\right]_{r_A}^{r_B},\nonumber
\eea
or alternatively
\begin{eqnarray}\label{shifthtBL}
&\Phi_{kj}\simeq\mp\frac{\Delta m_{kj}^2}{2E}\Delta r\,\times\\
&\left[1+\frac{m^2}{r_Br_A}\left\{q+\frac{b^2}{2m^2}+(r_B+r_A)\frac{2q(b^2+m^2)+b^2}{2mr_Br_A}\right\}\right],\nonumber
\end{eqnarray}
where $\Delta r\equiv r_B-r_A$ and the second and higher order terms in $q$ as well as the weak field expansions $m/r\ll1$ have been neglected.
In the limiting case of vanishing quadrupole parameter, Eq. (\ref{shifthtBL}) reproduces previous results developed in the literature~\cite{sol4,sol5} for the Schwarzchild space-time.

It is also useful to replace the parameters $m$ and $q$ by the total mass and quadrupole moment\footnote{According to Geroch-Hansen definitions of the multipole moments, we can match the quadrupole moment of Hartle and Thorne with the $q$-metric by $Q_{GH}=Q_{q}= - Q_{HT}=Q_{GL}-2J^2/M$. Hence, positive $Q_{HT}=Q>0$ is for oblate objects and vice versa. $Q_{GL}$ is the quadrupole moment defined by ~\cite{sol4}.} $Q_{q}$ \cite{2018RSOS....570826F}
\be
M=m(1+q)\,,\qquad
Q_{q}=-\frac{2}{3}m^3q\,.
\ee
For instance, when $b=0$, Eq. (\ref{shifthtBL}) reads
\be
\label{shifthtBL2}
\Phi_{kj}\equiv\Phi_{kj}^{\rm(m)}+\Phi_{kj}^{\rm(q)}\ ,
\ee
where the monopole ${\rm(m)}$ and quadrupole ${\rm(q)}$ moments are
\begin{eqnarray}
\label{shifthtBL3}
&\Phi_{kj}^{\rm (m)}&=\mp\frac{\Delta m_{kj}^2}{2E}\Delta r\ , \\
&\Phi_{kj}^{\rm (q)}&=\,\Phi_{kj}^{\rm(m)}\frac{ M^2}{r_Br_A}\left[1+\frac{M(r_B+r_A)}{r_Br_A}\right]\left(\frac{3}{2}\frac{Q_{q}}{M^3}\right)\nonumber \ .
\end{eqnarray}
The monopole term is the dominant one, due to the large distance between the source and detector. However, describing the background gravitational field simply by using the spherically symmetric Schwarzschild solution is not satisfactory in most situations.
In fact, astrophysical sources are expected to be rotating as well endowed with shape deformations leading to effects which cannot be neglected in general\footnote{
The modification to the phase shift induced by space-time rotation has been already taken into account in Ref. \cite{wudka}.}.

It is worth pointing out that  Eq. (\ref{shifthtBL3}) and following deal with the oscillation baseline $\Delta r$ and the asymptotic neutrino energy $E$. To make a comparison with the experiments, these quantities have to be expressed as measured by a locally inertial observer at rest with the oscillation experiment. In general for both source and observer at rest with respect to the reference frame, the proper oscillation baseline is given by $\Delta r=c\int\sqrt{g_{00}(x^\mu_{obs})}dt$, while the observed neutrino energy $E_{obs}=E_{em}\sqrt{g_{00}(x^\mu_{em})/g_{00}(x^\mu_{obs})}$ relates to the emitted one $E_{em}$. For distant observers we have $g_{00}(x^\mu_{em})=e^{2\psi}$ and $g_{00}(x^\mu_{obs})\approx1$, therefore the experimental setup measures effectively $\Delta r=c\Delta t$ and $E_{obs}=e^\psi E_{em}$; for observers close to the source of neutrinos, as we are going to consider in the next sections, we have $g_{00}(x^\mu_{em})\approx g_{00}(x^\mu_{obs})=e^{2\psi}$, therefore the experimental setup measures effectively $\Delta r=c\int e^\psi dt$ and $E_{obs}= E_{em}$.
In the following, the above correction will be included in the definitions of the baseline and $E$, if not otherwise specified.


\section{Numerical constraints}
\label{excons}

It is possible now to draw some considerations by using experimental data. In particular, we can split two different data surveys in which it is possible to take data points. The first set is based on cosmological constraints, whereas the second by data coming from reactors and nuclear physics in general. Let us explore in detail both the possibilities below.
\vspace{0.2cm}

{\bf \emph{Cosmological constraints.}} The current limit on the sum of the neutrino masses has been obtained from the analysis of the cosmic microwave background anisotropy combined with the galaxy redshift surveys and other data and has been set to $\sum_i m_{\nu_{i}} \le 0.7~\rm{eV}$ \cite{WMAP}.
On the other hand, Big Bang nucleonsynthesis gives constraints on the total number of neutrinos, including possible sterile neutrinos which do not interact and are produced only by mixing. The number is currently $1.7 \le N_{\nu} \le 4.3$ at 95\% of confidence level \cite{WMAP}. More recent results seem to indicate tighter limits over neutrino masses. This has been found by the Planck satellite where analyses made by combining more data sets  constrain the effective extra relativistic degrees of freedom to be compatible with the standard cosmological model predictions, with neutrino masses constrained by $\sum_i m_{\nu_i}\leq 0.12$ \cite{PLANCK2018}.

\vspace{0.2cm}

{\bf \emph{``Reactor-like" data.}} Long-baseline reactors and low-energy Solar neutrino experiments, in which the matter effects are subdominant compared to vacuum oscillations, are specially suited to estimate the parameter phase space for the mass eigenstates $1$ and $2$ \cite[see, e.g.,][]{2013arXiv1303.4667T}.

Recently, a new global fit of neutrino oscillation parameters, within the simplest three-neutrino, obtained by including
\begin{itemize}
    \item[{\bf 1)}] {new long-baseline disappearance and appearance data involving the antineutrino channel in T2K, and the $\nu_\mu$-disappearance and $\nu_e$-appearance data from NO$\nu$A,}
    \item[{\bf 2)}] {reactor data such as from the $\bar{\nu}_e$-disappearance spectrum of Daya Bay, the prompt reactor spectra from RENO, and the Double Chooz event energy spectrum,}
    \item[{\bf 3)}] atmospheric neutrino data from the IceCube DeepCore and ANTARES neutrino telescopes, and from Super-Kamiokande, and
    \item[{\bf 4)}] solar oscillation spectrum from Super-Kamiokande,
\end{itemize}
has established that  $\Delta\tilde{m}^2_{21}=7.55^{+0.20}_{-0.16}\times10^{-5}$~eV$^2$ within the normal ordering picture \cite[see][for details]{DESALAS2018633}.

A part from the possible sterile neutrinos, cosmological constraints are in agreement with the current paradigm purporting the existence of three different neutrino mass eigenstates.
Global analyses of neutrino and antineutrino experiments seem to favor the normal hierarchy of the \emph{three mass eigenstates}, namely ($m_3\gg m_2>m_1$) that we pursue throughout this work  \cite{2017PhRvD..95i6014C,DESALAS2018633,2017PhRvD..95i6014Cbis}.

In the following we consider the experimental values from reactos and low energy solar neutrino experiments.

\subsection{Computation of numerical bounds in the weak field regime}

With the numerical limits imposed by the previous discussion above, we can now compute the phase shift on the surfaces of:
\begin{itemize}
    \item[{\bf a.}] the Earth, from nuclear reactors, that can be built up with current technology, and
    \item[{\bf b.}] the Sun, detected from an hypothetical neutrino detector in its proximity.
\end{itemize}

Let us indicate the mass and the radius of the above astronomical objects with general labels $M_\star$ and $R_\star$, respectively.
Their quadrupole moments can be expressed as $Q_\star=-J_2^\star M_\star R^2_\star$, where $J_2^\star$ is the dimensionless quadrupole moment.
On the surface of the above the astronomical objects we can use the following approximations $r_B+r_A\approx2 R_\star$, $r_B r_A\approx R_\star^2$, and $r_B-r_A\simeq d$, where $d$ is the oscillation baseline.
Hence, after cumbersome algebra and replacing the above approximations in Eq.~(\ref{shifthtBL2}) we can compute the effect of $Q_\star$ to the shift phase
\bea
\label{shifthtBL2bis}
\Phi_{21}=\frac{\Delta m_{21}^2}{2E}d
\left[1-\frac{3}{2}J_2^\star\left(1+\frac{2 M_\star}{R_\star}\right)\right]\ ,
\eea
where the sign ``$\mp$'' has been ignored since it does not affect the following analysis. Moving from the assumption that the experiments measure the phase shift affected by the gravitational effects $\Delta\tilde{m}^2_{21}$, from Eq.~(\ref{shifthtBL2bis}) it is immediately clear that this mass difference is essentially given by
\bea
\label{massexp}
\Delta\tilde{m}^2_{21}=\Delta m_{21}^2
\left[1-\frac{3}{2}J_2^\star\left(1+\frac{2 M_\star}{R_\star}\right)\right]\ ,
\eea
where $\Delta m_{12}^2$ is the real mass difference between the neutrino eigenstates 1 and 2.

\begin{table*}
\setlength{\tabcolsep}{0.75em}
\renewcommand{\arraystretch}{1.0}
\centering
\caption{The estimate of $\Delta m_{21}^2$ without the effects of the quadrupole moment for the Earth ($\oplus$) and the Sun ($\odot$). Columns list, respectively: the dimensionless quadrupole moment, the mass and the radius of the object, the inferred value of $\Delta m_{21}^2$, and the percent departure from the experimental value $\Delta\tilde{m}^2_{21}$.}
\small
\begin{tabular}{cccccc}
\hline\hline
Object
		& $J_2^\star$
		& $M_\star$
		& $R_\star$
		& $\Delta m^2_{21}$
		& $1- \Delta m^2_{21}/\Delta \tilde{m}^2_{21}$\\
		&
		& (m)
		& (m)
		& ($10^{-5}$~eV$^2$)
		& (\%)\\
\hline
          $\oplus$
        & $1.0826\times10^{-3}$
		& $4.4350\times10^{-3}$
		& $6.3781\times10^6$
		& $7.56^{+0.20}_{-0.16}$
		& $-0.2$\\
          $\odot$
        & $2.1106\times10^{-7}$
		& $1.4771\times10^{3}$
		& $6.9551\times10^8$
		& $7.55^{+0.20}_{-0.16}$
		& $\approx0$\\
\hline
\end{tabular}
\label{tab:no0}
\end{table*}

Using the the actual values of $J_2^\star$, $M_\star$, and $R_\star$ for the Earth \cite{2014AmJPh..82..769T} and the Sun \cite{rozelot}, employing the above value of $\Delta\tilde{m}^2_{21}$ given by \cite{DESALAS2018633}, and by reverting Eq.~(\ref{massexp}) one obtains the correction induced by the quadrupole contribution on $\Delta m_{12}^2$ and the percent departure from the experimental value (see Tab.~\ref{tab:no0}).
The inferred values of $\Delta m_{12}^2$ are indistinguishable from the experimental one and so we definitively conclude that the effects due to the quadrupole moment turn out to be negligibly small, for both the Earth and Sun, in agreement with previous results.

For completeness, immediately after the study of the Earth and Sun, one can investigate cosmological sources for neutrino phase shift. To do so, the simplest idea is to take standard candles, widely used in observational cosmology, and to measure neutrino oscillations from their sources. However, the aforementioned considerations and limits suggest an intriguing fact: \emph{standard candles, such as supernovae Ia, cannot be treated at this step}. In fact, the Solar System regime involves weak gravitational fields while, in the framework of supernovae Ia, neutrinos are produced in the nuclear processes taking place in the WD. In a similar way, in the case of supernovae II, what produces neutrinos is the newly-born NS at the center of their explosion.

This clearly represents a limitation, since supernovae are objects of great interest in cosmology, whose physical properties are well established by observations. A possible hint is to consider supernova explosions as indications for investigating neutrinos from WDs or NSs for example, in the strong gravitational field. Hence, motivated by this fact, in the next subsection, we investigate how the quadrupole moment affects the value of $\Delta \tilde{m}_{12}^2$ in the case of rotating WDs and NSs, i.e. two astrophysical configurations in which the neutrino phase shift may produce more relevant  results.

\subsection{Rotating white dwarfs and neutron stars}
\label{wdns}

\begin{figure*}
\centering
\begin{tabular}{lr}
\includegraphics[width=3.3in]{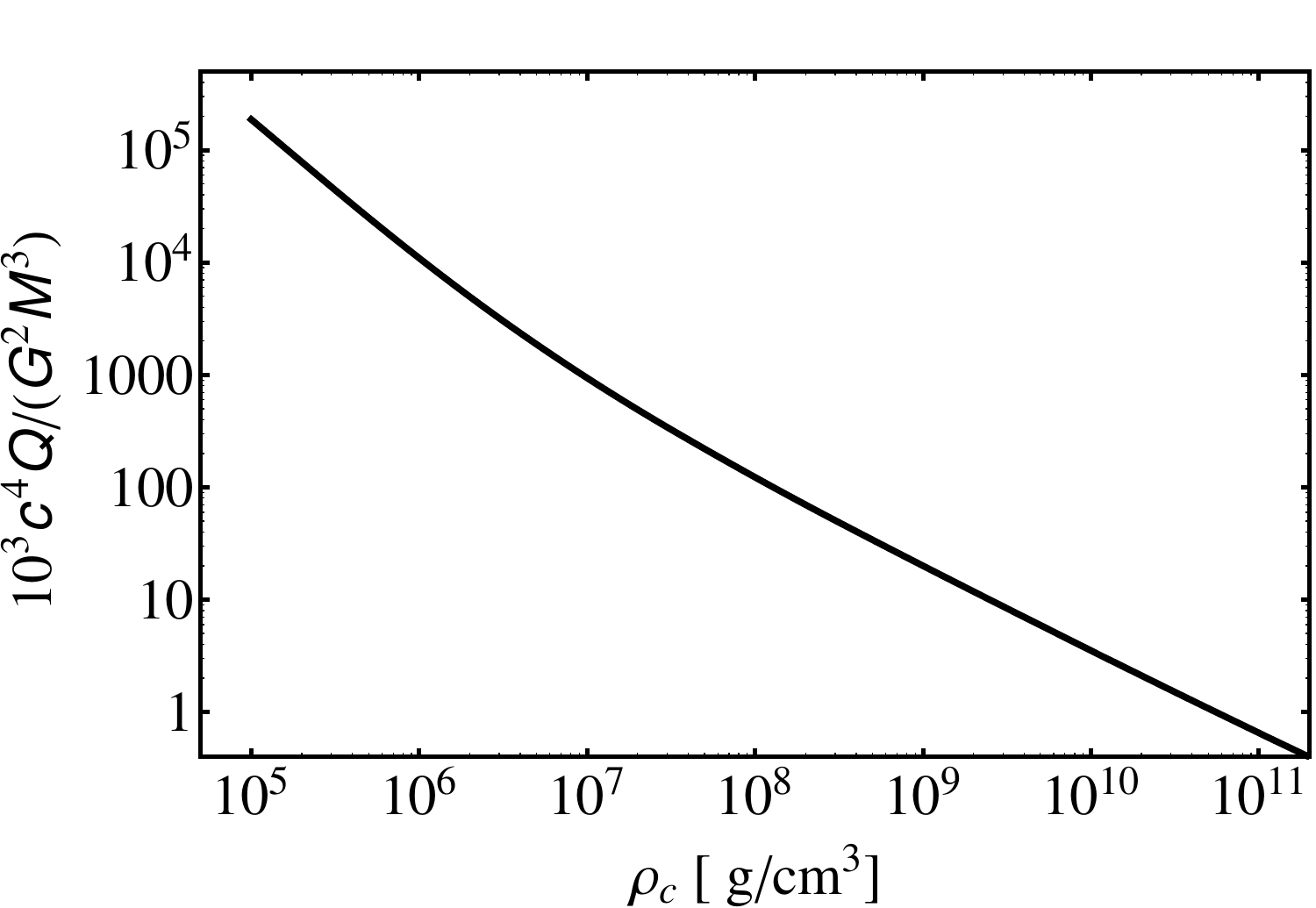} & \includegraphics[width=3.3in]{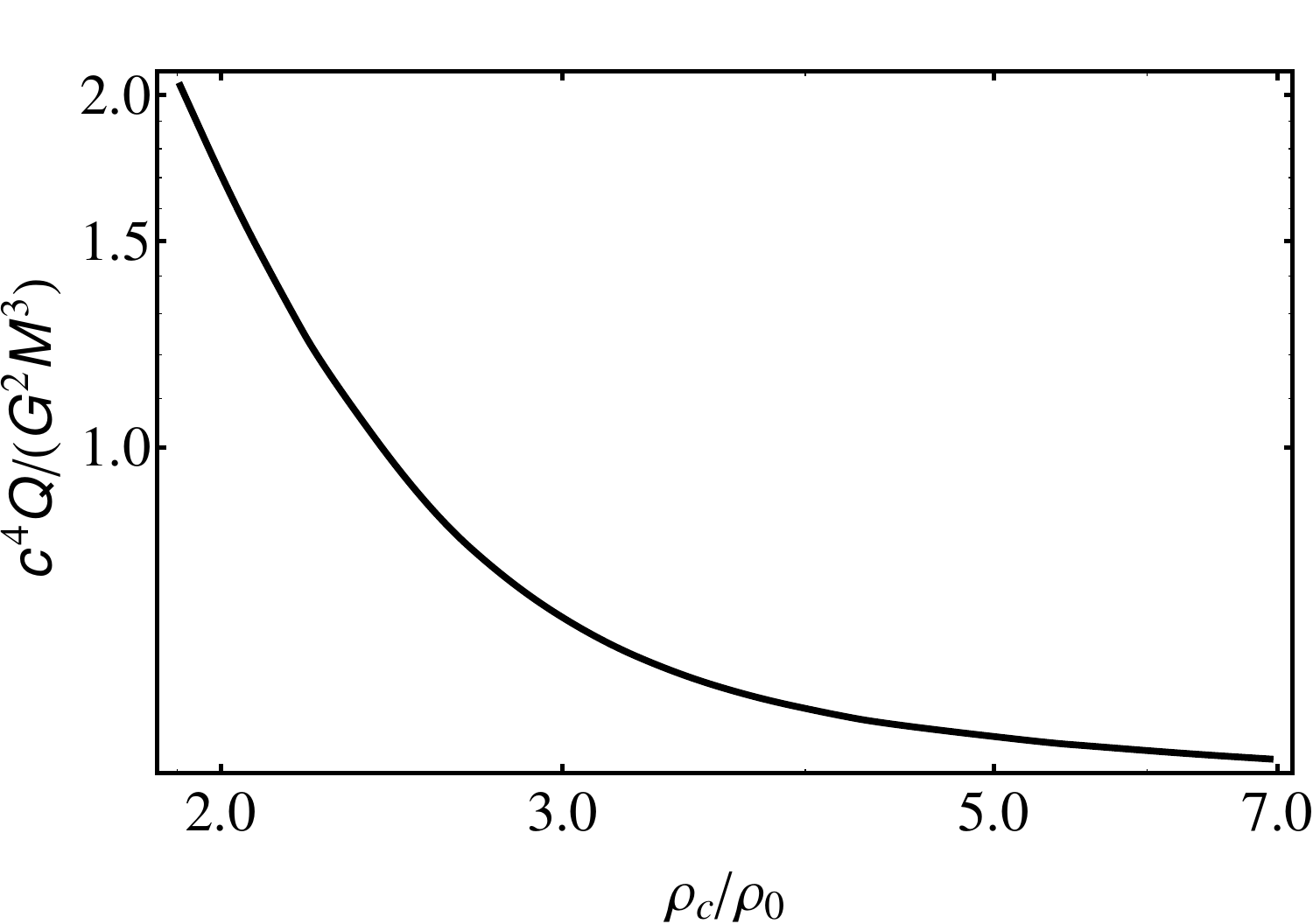}
\end{tabular}
\caption{ The mass quadrupole moment $Q$ over the total mass cubed $M^3$ versus central density $\rho_c$.  Left panel: Maximally rotating WDs. Right panel: Maximally rotating NSs, where $\rho_0$ is the nuclear density. }\label{fig:QM_vs_rho}
\end{figure*}

\begin{figure*}
\centering
\begin{tabular}{lr}
\includegraphics[width=3.3in]{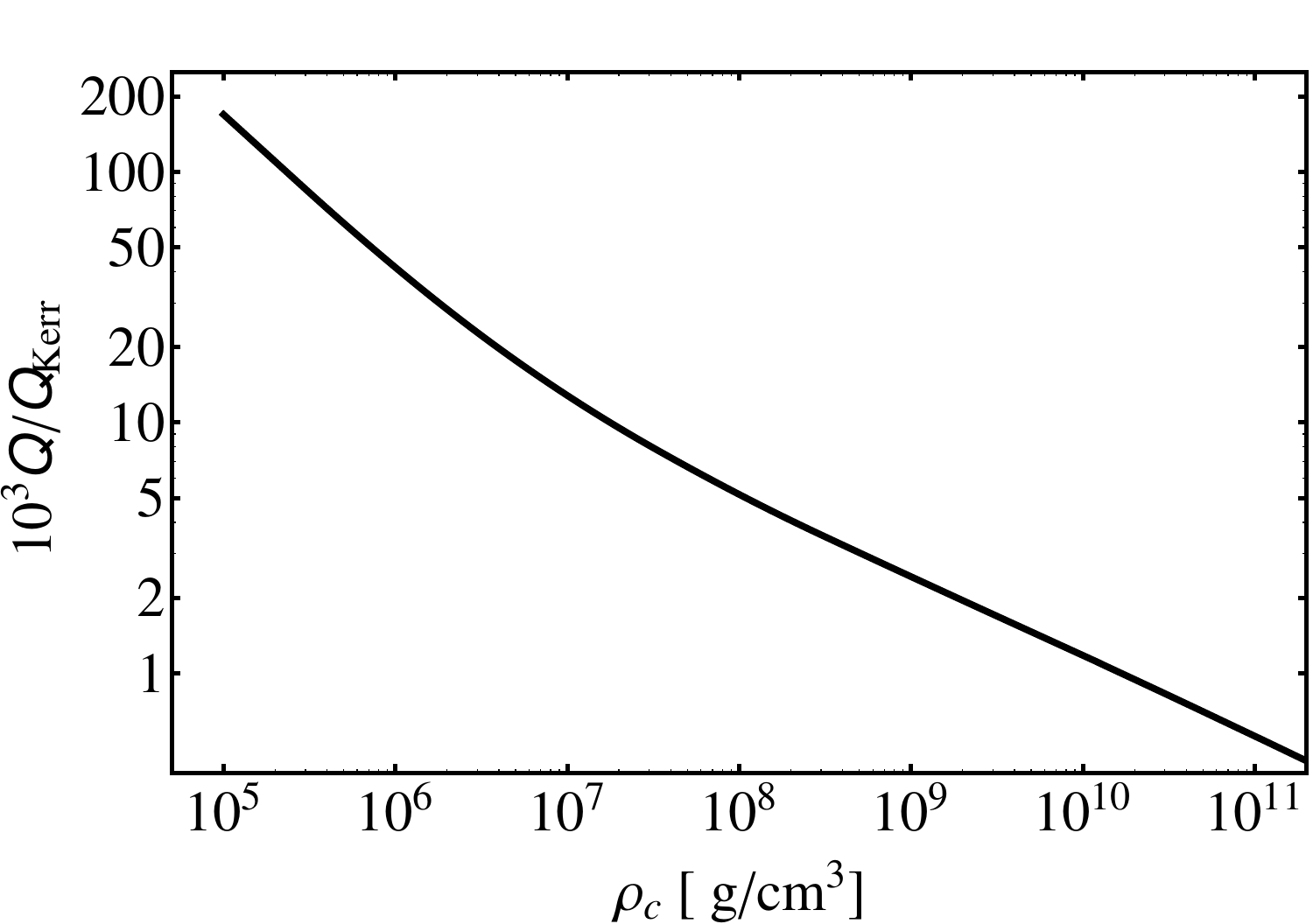} & \includegraphics[width=3.3in]{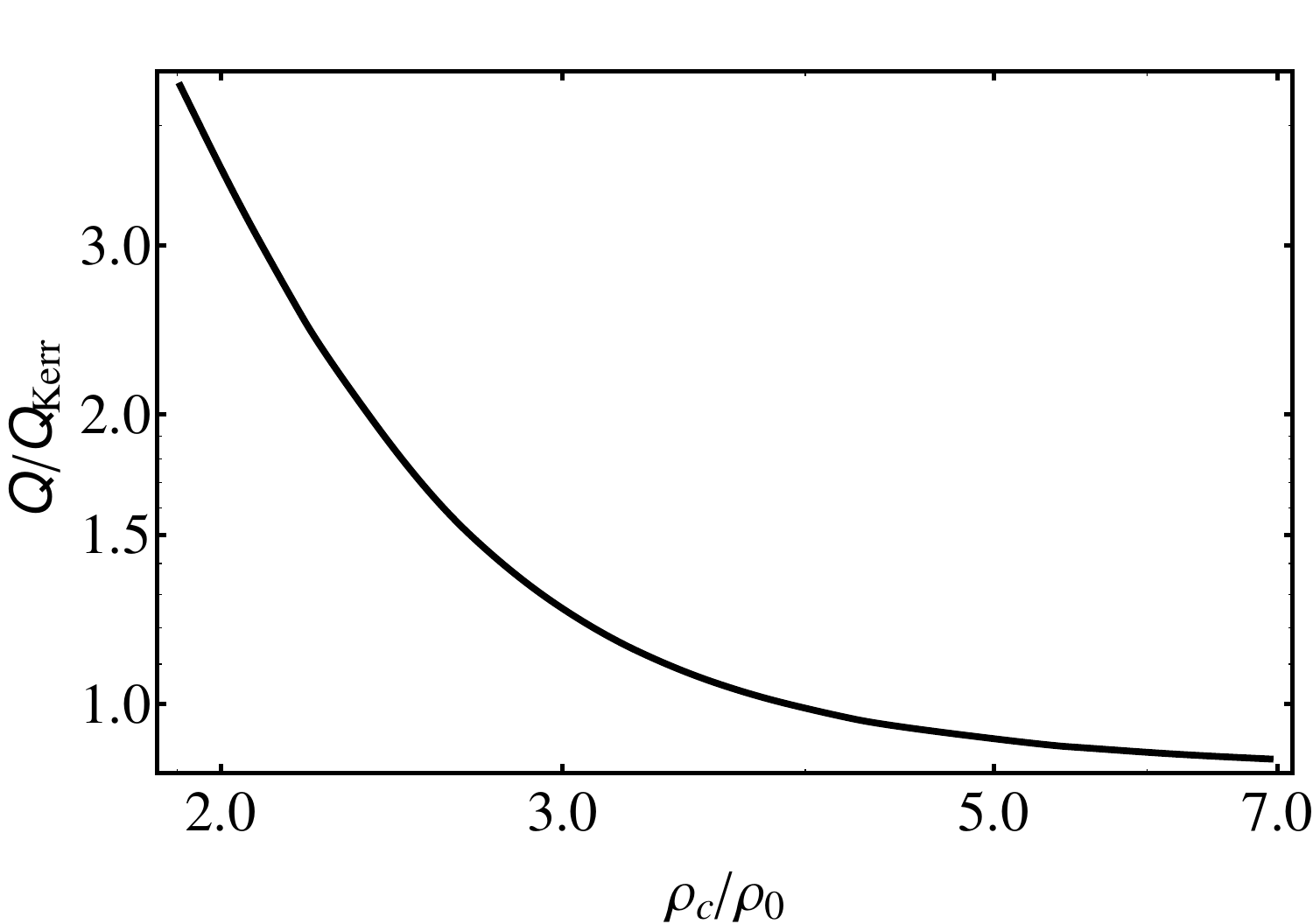}
\end{tabular}
\caption{ The mass quadrupole moment $Q$ over the Kerr quadrupole moment $Q_{\rm Kerr}=J^2/(c^2M)$ versus central density $\rho_c$.  Left panel: Maximally rotating WDs. Right panel: Maximally rotating NSs, where $\rho_0$ is the nuclear density. }\label{fig:Q_vs_rho}
\end{figure*}

\begin{table*}
\setlength{\tabcolsep}{0.75em}
\renewcommand{\arraystretch}{1.0}
\centering
\caption{The effects of the quadrupole moment (for {\it q}- and HT metrics) and the dipole moment (for HT metric only), for both WDs and NSs, on $\Delta m_{21}^2$. All the parameters have been estimated for maximally-, $0.1\times$maximally-, and  $0.01\times$maximally-rotating compact objects. In order, the columns list: the central density $\rho_c$ in g/cm$^3$ and in units of the nuclear density $\rho_0=2.7\times10^{14}$ g/cm$^3$, the mass $M$ in solar masses $M_\odot$=1.47 km , the radius $R$, and the quadrupole $Q$ and the dipole $J$ moments of the compact object; the following columns summarize the mass difference $\Delta\tilde{m}^2_{21,i}$ and the difference $\delta\Delta m^2_{21,i}\equiv \Delta\tilde{m}^2_{21,i}-\Delta m^2_{21}$ for the {\it q}-metric from Eq.~(\ref{massexpq}) and for the HT metric from Eq.~(\ref{massexpHT}).}
\small

\begin{tabular}{lccccccccc}
\hline\hline
\multicolumn{10}{c}{Maximally-rotating configurations}\\
\cline{1-10}
		& $\rho_c$
		& $M$
		& $R$
		& $Q$
		& $J$
		& $\Delta\tilde{m}^2_{21,q}$
		& $\Delta\tilde{m}^2_{21,HT}$
		& $\delta\Delta m^2_{21,q}$
		& $\delta\Delta m^2_{21,HT}$\\

		& (g/cm$^3$)
		& (M$_\odot$)
		& ($10^3$~m)
		& ($10^9$~m$^3$)
		& ($10^4$~m$^2$)
		& ($10^{-5}$~eV$^2$)
		& ($10^{-5}$~eV$^2$)
		& ($10^{-5}$~eV$^2$)
		& ($10^{-5}$~eV$^2$)\\
\hline
		& $10^5$
		& 0.18
		& 18304.5
		& 3352780.0
		& 208.54
		& $7.13_{-0.15}^{+0.18}$
		& $7.19_{-0.15}^{+0.19}$
		& -0.43
		& -0.37\\

WD		& $10^6$
		& 0.47
		& 12260.6
		& 3727340.0
		& 728.23
		& $7.16_{-0.15}^{+0.19}$
		& $7.23_{-0.15}^{+0.19}$
		& -0.40
		& -0.34\\

		& $10^8$
		& 1.33
		& 4859.3
		& 912628.0
		& 1753.62
		& $7.34_{-0.16}^{+0.19}$
		& $7.37_{-0.16}^{+0.20}$
		& -0.22
		& -0.19\\

\hline
		& $1.67\rho_0$
		& 1.07
		& 13.61
		& 13.48
		& 151.09
		& $6.88_{-0.15}^{+0.18}$
		& $7.03_{-0.15}^{+0.19}$
		& -0.68
		& -0.53\\

NS		& $1.91\rho_0$
		& 1.45
		& 13.86
		& 19.92
		& 290.24
		& $6.78_{-0.14}^{+0.18}$
		& $6.99_{-0.15}^{+0.19}$
		& -0.78
		& -0.57\\

		& $2.55\rho_0$
		& 2.18
		& 14.00
		& 29.50
		& 717.26
		& $6.63_{-0.14}^{+0.18}$
		& $7.06_{-0.15}^{+0.19}$
		& -0.94
		& -0.50\\

\hline\hline
\multicolumn{10}{c}{$0.1\times$Maximally-rotating configurations}\\
\cline{1-10}
		& $10^5$
		& 0.15
		& 16383.0
		& 33527.8
		& 20.85
		& $7.56_{-0.16}^{+0.20}$
		& $7.56_{-0.16}^{+0.20}$
		& $<-0.001$
		& $<-0.01$\\

WD		& $10^6$
		& 0.40
		& 10971.9
		& 37273.4
		& 72.82
		& $7.56_{-0.16}^{+0.20}$
		& $7.56_{-0.16}^{+0.20}$
		& $<-0.001$
		& $<-0.01$\\

		& $10^8$
		& 1.19
		& 4332.8
		& 9126.3
		& 175.36
		& $7.56_{-0.16}^{+0.20}$
		& $7.56_{-0.16}^{+0.20}$
		& $<-0.01$
		& $<-0.01$\\
\hline
		& $1.67\rho_0$
		& 0.91
		& 12.70
		& 0.13
		& 15.11
		& $7.55_{-0.16}^{+0.20}$
		& $7.56_{-0.16}^{+0.20}$
		& $<-0.01$
		& $<-0.01$\\
		
NS		& $1.91\rho_0$
		& 1.27
		& 13.13
		& 0.20
		& 29.02
		& $7.55_{-0.16}^{+0.20}$
		& $7.56_{-0.16}^{+0.20}$
		& $<-0.01$
		& $<-0.01$\\

		& $2.55\rho_0$
		& 2.00
		& 13.66
		& 0.29
		& 71.73
		& $7.55_{-0.16}^{+0.20}$
		& $7.56_{-0.16}^{+0.20}$
		& $-0.0103$
		& $<-0.01$\\
\hline\hline
\multicolumn{10}{c}{$0.01\times$Maximally-rotating configurations}\\
\cline{1-10}
		& $10^5$
		& 0.15
		& 16363.8
		& 335.28
		& 2.09
		& $7.56_{-0.16}^{+0.20}$
		& $7.56_{-0.16}^{+0.20}$
		& $<-0.01$
		& $<-0.01$\\

WD		& $10^6$
		& 0.40
		& 10959.0
		& 372.73
		& 7.28
		& $7.56_{-0.16}^{+0.20}$
		& $7.56_{-0.16}^{+0.20}$
		& $<-0.01$
		& $<-0.01$\\

		& $10^8$
		& 1.18
		& 4327.5
		& 91.26
		& 17.54
		& $7.56_{-0.16}^{+0.20}$
		& $7.56_{-0.16}^{+0.20}$
		& $<-0.01$
		& $<-0.01$\\

		\hline
		& $1.67\rho_0$
		& 0.91
		& 12.69
		& 0.0013
		& 1.51
		& $7.56_{-0.16}^{+0.20}$
		& $7.56_{-0.16}^{+0.20}$
		& $<-0.01$
		& $<-0.01$\\

NS		& $1.91\rho_0$
		& 1.26
		& 13.12
		& 0.002
		& 2.90
		& $7.56_{-0.16}^{+0.20}$
		& $7.56_{-0.16}^{+0.20}$
		& $<-0.01$
		& $<-0.01$\\

		& $2.55\rho_0$
		& 1.99
		& 13.65
		& 0.003
		& 7.17
		& $7.56_{-0.16}^{+0.20}$
		& $7.56_{-0.16}^{+0.20}$
		& $<-0.01$
		& $<-0.01$\\

\hline
\end{tabular}

\label{tab:no1}
\end{table*}

\begin{table*}
\setlength{\tabcolsep}{1.0em}
\renewcommand{\arraystretch}{1.2}
\caption{Principal parameters and results involved in our landscapes. The table summarizes the effects of the quadrupole moment for the {\it q}- and the HT metrics in the regimes of weak and strong gravity on $\Delta\tilde{m}^2_{21}$. The expected values are reported in view of possible future experiments.}

\begin{tabular}{lcl}
\hline\hline
\multicolumn{3}{c}{{\bf Regime of weak gravity}}\\
\cline{1-3}
		& $Main\,\,\,parameters$ & $Outstanding\,\,\,results$ \\

\hline
		&  $M\in[M_\oplus;M_\odot]$ & Results using HT and $q$ metrics are indistinguishable.\\

{\bf \emph{Solar System}}		&  Quadrupole moment: $Q_\star$  & Spherical symmetric effects are the unique to be measured.\\

		& $\Phi_{kj}^{\rm HT,(m)}\gg\Phi_{kj}^{\rm HT,(d)}$ & $\delta\Delta m^2_{21}/\Delta \tilde{m}^2_{21}$ currently indistinguishable from the flat case\\

\hline\hline
\multicolumn{3}{c}{{\bf Regime of strong gravity}}\\
\cline{1-3}
		& $Main\,\,\,parameters$ & $Outstanding\,\,\,results$ \\

\hline
		& $M\in[0.15;2.18]\,M_\odot$ & Deviations are observable only for  maximally-rotating compact objects.\\

{\bf \emph{NSs} $\,\&\,$ \emph{WDs}}		& $Q\in[10^5;10^{16}]\,{\rm m}^3$& Results using HT and $q$ metrics are indistinguishable.\\

		& NSs: $\Phi_{kj}^{\rm HT,(d)}\gg\Phi_{kj}^{\rm HT,(q)}$ & Instrument sensibility  can probe $1\sigma$ deviations from spherical symmetry. \\
	& WDs: $\Phi_{kj}^{\rm HT,(d)}\ll\Phi_{kj}^{\rm HT,(q)}$ &  \\

\hline\hline
\end{tabular}

\label{due}
\end{table*}

The computation of the basic parameters of rigidly rotating WDs and NSs is not straightforward, as it may seem at first glance. It is related to the fact that unlike in Newtonian gravity where the field equations, the equations of motion (hydrostatic equilibrium and the mass balance equations) are given separately, in GR all these equations are contained in the Einstein gravitational field equations\footnote{One should derive them directly from the field equations and solve them for matter, preliminary adopting an equation of state (EoS) and external vacuum. On the matter-vacuum interface one has to perform the matching between the metric functions, finding out the integration constants.}. The procedures to follow are well-known in the literature \cite{shapirobook,haenselbook} and for static objects the field equations reduce to the Tolman-Oppenheimer-Volkoff equations \cite{tolman39,oppenheimer39}.

Due to rotation the Tolman-Oppenheimer-Volkoff equations i.e. the mass balance, the hydrostatic equilibrium and gravitational potential equations will be modified. The solutions of those equations with a chosen EoS will yield the parameters of rotating objects such as angular momentum, quadrupole moment etc. In our computations we make use of the Hartle formalism, which allows one to construct and investigate the equilibrium configurations of slowly rotating stellar objects \cite{1967ApJ...150.1005H,1968ApJ...153..807H}. For qualitative rough analyzes one may employ the Hartle formalism at mass-shedding limit, though for quantitative analyzes one should use full GR equations \cite{stergioulas,berti2005}.

It is well know that unlike magnetic field or anisotropic pressure, rotation is the main contributor to the deformation of compact objects such as WDs and NSs \cite{2014NuPhA.921...33B,2017IJMPS..4560025T, 2019MNRAS.487..812B}. Here by exploiting the HT formalism \cite{1967ApJ...150.1005H, 1968ApJ...153..807H, 2016EJPh...37f5602B} the rotating equilibrium configurations of WDs \cite{2013ApJ...762..117B,2017MNRAS.464.4349B} and NSs \cite{2014NuPhA.921...33B,NSbook,2018mgm..conf.3433B} are constructed and  the mass quadrupole moment as a function of central density for maximally rotating stars (see Fig.~\ref{fig:QM_vs_rho}) are calculated and compared with the Kerr quadrupole moment which is related to the angular momentum of the source (see Fig.~\ref{fig:Q_vs_rho}).
The EoS of NSs is taken from \cite{2012NuPhA.883....1B}, where all fundamental interactions are taken into account. We adopted the so-called NL3 model, well-known in the compact object literature, in the EoS for NS, which was derived in the frame of the relativistic mean field theory with nonlinear parametrization set (see \cite{Reinhard}, \cite{Sharma}, \cite{Lalazissis} for further details). This EoS is one of many stiff equations of state which are in accordance with the observational constraints on NSs \cite{2014NuPhA.921...33B}.
The  degenerate electron gas EoS was employed for the WD matter \cite{chandrasekhar31,2015mgm..conf.2468B} as it is the simplest EoS.

For our purposes, in Fig.~\ref{fig:QM_vs_rho} the ratio of the mass quadrupole moment $Q$ to the total mass cubed is given as a function of the central density in physical units. Here $Q$ is rotationally induced. The value of $Q/M^3$ is larger in WDs (left panel) than in NSs (right panel) this means that rotation deforms strongly objects with a soft EoS, whereas the EoS of a NS, considered here, is stiff.

On the contrary, Fig.~\ref{fig:Q_vs_rho} shows the ratio of the mass quadrupole moment to the Kerr quadrupole moment as a function of the central density for maximally rotating WDs (left panel) and NSs (right panel). As one can see, the quadrupole moment contribution is larger than that due to spin (angular momentum) in the case of WDs. However, in the case of NSs,  the spin contribution may be larger than that due to deformation.

It is worth stressing that, even though rotation (and correspondingly angular momentum) induces deformation of stellar objects, its effect on the phase shift of neutrino oscillations becomes significant only in NSs. Therefore we confirm the previous results obtained in \cite{sol4},
where the phase shift was computed employing the HT space-time \cite{1967ApJ...150.1005H,bglq}. Although in Figs~\ref{fig:QM_vs_rho}-\ref{fig:Q_vs_rho} we have maximally rotating objects, their slow-rotation limit will display a similar behavior though the value of $Q$ will be much lower. Thus, for both WDs \cite{2013ApJ...762..117B,2017MNRAS.464.4349B} and NSs \cite{2014NuPhA.921...33B,NSbook,2018mgm..conf.3433B}, we select three different mass-radius configurations computed for the following three cases:
\begin{itemize}
    \item[{\bf A.}]  maximally-rotating objects;
    \item[{\bf B.}] $0.1\times$maximally-rotating objects;
    \item[{\bf C.}] $0.01\times$maximally-rotating objects.
\end{itemize}
For both NSs and WDs, we compute the quadrupole-induced mass difference $\Delta\tilde{m}^2_{21}$ from the real value $\Delta m_{21}^2$ inferred previously, splitting, for WDs and NSs, the phase for the $q$-metric by
\bea
\label{deltaphistrongreg}
\Phi_{kj}^{\rm q}\equiv\Phi_{kj}^{\rm q,(m)}+\Phi_{kj}^{\rm q,(q)}\,,
\eea
where $\Phi_{kj}^{\rm q,(m)}$ is given by the first relation of Eqs. (\ref{shifthtBL3}) while
\begin{eqnarray}
\label{eq:psqsf2}
\Phi_{kj}^{\rm q,(q)}\equiv&\frac{3 Q_q \Phi_{kj}^{\rm q,(m)} }{M^2\Delta r} \left\{\log\left[\frac{(r_A-M)(r_B-2 M)}{(r_A-2 M)(r_B-M)}\right]+\right.\\
&\left.\frac{r_A}{2M} \log\left[\frac{r_A(r_A-2 M)}{(M-r_A)^2}\right]-\frac{r_B}{2M} \log\left[\frac{r_B (r_B-2 M)}{(M-r_B)^2}\right]\right\}.\nonumber
\end{eqnarray}

\noindent Note that, in the weak field regime,  Eq. (\ref{eq:psqsf2}) reduces to the corresponding one given by Eq.~(\ref{shifthtBL3}).

\noindent Since $q$-metric describes a static and deformed astrophysical object, we check the above results by computing the phase shift within the HT metric in the strong field regime:\footnote{In the HT space-time definition the quadrupole $Q$ moment is defined positive (negative) for oblate (prolate) objects, while for the $q$-metric the viceversa holds.}
\bea
\label{deltaphistrongregHT}
\Phi_{kj}^{\rm HT}\simeq\Phi_{kj}^{\rm HT,(m)}+\Phi_{kj}^{\rm HT,(d)}+\Phi_{kj}^{\rm HT,(q)}\,,
\eea
where again $\Phi_{kj}^{\rm HT,(m)}$ is given by the first relation of Eqs.~(\ref{shifthtBL3}) and the dipole ${\rm (d)}$ and quadrupole terms are
\begin{eqnarray}
\Phi_{kj}^{\rm HT,(d)}=&-\frac{J^2}{M}\left[\frac{\Phi_{kj}^{\rm HT,(q)}}{Q} + \frac{\Phi_{kj}^{\rm HT,(m)}}{2M}\frac{M \left(r_A+r_B\right)+ r_A r_B}{r_A^2 r_B^2}\right],\nonumber\\
\label{HTJQ2}
\Phi_{kj}^{\rm HT,(q)}=&\frac{15 Q \Phi_{kj}^{\rm HT,(m)}}{8 M^3} \times
\left[1+\frac{r_A}{\Delta r} g_A-\frac{r_B}{\Delta r}g_B\right]\ ,
\end{eqnarray}
where
\begin{equation}
g_i\equiv \left(1-\frac{r_i}{2 M}\right) \log \left(1-\frac{2 M}{r_i}\right)\,.
\end{equation}
Note that in the weak field also  Eq.~(\ref{HTJQ2}) reduces to the one in \cite{sol4} by replacing $Q=Q_{HT}=2J^2/M-Q_{GL}$.

As one may notice, unlike Eq.~(\ref{shifthtBL3}), the corrections to the monopole term in Eqs.~(\ref{deltaphistrongreg}--\ref{HTJQ2}) display the dependence upon the neutrino oscillation baseline. To get rid of it, we study the neutrino oscillation occurring at the surface of the compact object by setting $r_B=r_A+\Delta r$, where $r_A\equiv R$ is the compact object radius averaged between the equatorial and the polar radii, defined as $R=(R_{\rm pol}+2R_{\rm eq})/3$, and expand Eqs.~(\ref{deltaphistrongreg}--\ref{HTJQ2}) around $\Delta r\approx0$ at the lowest order.
Finally, like in Eq.~(\ref{massexp}), we get the measured mass difference in the strong field regime for the {\it q}-metric
\bea
\label{massexpq}
\Delta\tilde{m}^2_{21,q}=\Delta m_{21}^2
\left\{1-\frac{3 Q_q}{2 M^3} \log\left[\frac{R \left(R-2 M\right)}{\left(M-R\right)^2}\right]
\right\}\,,
\eea
and for the HT metric
\begin{eqnarray}
\label{massexpHT}
\Delta\tilde{m}^2_{21,HT}=
&\Delta m_{21}^2\left\{1-\frac{J^2\left(R+2M\right)}{2M^2 R^3}+\frac{15}{8 M^3}\left(Q-\frac{J^2}{M}\right)\times\right.\nonumber\\
&\left.\left[2-g(R)+\frac{R}{2 M}\log\left(1-\frac{2M}{R}\right)\right]\right\}\,.
\end{eqnarray}
The results are summarized in Tab.~\ref{tab:no1}.
As it is immediately clear, deviations with respect to the real value $\Delta m^2_{21}$ are observable only for (nearly) maximally-rotating WDs and NSs. In these cases the estimates $\Delta\tilde{m}^2_{21,q}$ and  $\Delta\tilde{m}^2_{21,HT}$ are both outside the experimental constraints on $\Delta m^2_{21}$.
In Tab.~\ref{tab:no1} we consider WDs and NSs having three distinct central densities in each case and different rotation rate. For maximally rotating configurations the corresponding parameters $M, R, Q, J$ are larger with respect to intermediate and slow rotation cases. This fact is due to the equilibrium conditions of the configurations possessing the same central densities \cite{1968ApJ...153..807H}.  The principal results are summarized in Tab.~\ref{due}.


\section{Theoretical discussion and experimental developments}
\label{gedanken}

We have obtained expressions describing the phase shift responsible for the neutrino oscillations in the case of deformed and static astrophysical objects described by the $q$-metric, both in weak and strong field regime. In the latter case, similar expressions have been derived  also by considering rotating objects described by the HT metric and compared to the case of the $q$-metric.
These expressions reduce to the Schwarzschild case for: a) vanishing quadrupole moment only, in the q-metric case, and b) for both vanishing dipole and quadrupole moments, in the HT metric case.
From the expressions of the phase shift obtained so far, we made use of the constraint from long-baseline reactor and low-energy solar neutrino experiments to get equivalent expressions for the mass difference $\Delta m^2_{21}$.

Our outcomes have shown that all the effects of rotation for spherically symmetric space-time are negligible, in agreement with previous estimates made with the use of Schwarzschild space-time. Deformations become quite relevant for massive objects, among all NSs and WDs. Under the simplest choice of describing such objects by means of $q$-metric, we have described a  static and deformed astrophysical object whose gravitational field generalizes the Schwarzschild metric, introducing a quadrupole term that becomes significant when the mass shedding limit (the maximum rotation rate) is taken into account.

As it is immediately clear, deviations with respect to the real value $\Delta m^2_{21}$, as well as from that from the Earth reactor experiments $\Delta\tilde{m}^2_{21}$, are observable only for (nearly) maximally-rotating WDs and NSs. In these cases the estimates from the $q$-metric $\Delta\tilde{m}^2_{21,q}$ and from the HT metric $\Delta\tilde{m}^2_{21,HT}$ are both outside the experimental constraints on $\Delta m^2_{21}$. This implies that a measurement of $\Delta\tilde{m}^2_{21}$ from such extreme astrophysical sources may give numerical constraints on their value of $J$ and $Q$ and, then, also on $R$ and $M$. Reducing the rotation rate implies that one can analyze different configurations with smaller quadrupole moment and angular momentum. In this way, by keeping the central density fixed, the mass and corresponding radius will be also decreased.

Although relevant, those results are jeopardized by the degeneracy which occurs in defining $r_B-r_A=d$ and in the choices on $r_B$ and $r_A$. The sensibility of current instruments are quite enough to probe deviations from the standard spherical case up to $1-$sigma confidence level. Over the past years, steady progresses in probing neutrino
masses have been carried forward by means of direct measurements of decay kinematics.

Several experiments have tried to measure net deviations from the case $m_\nu=0$. In particular, from the study of the shape of the $\beta$-decay spectrum near the end point energy, there is a very slight discrepancy between massless state and massive state plots. For the sake of clearness, the two plots exhibit sharp cut offs at the end point energy, although massive state plots smoothly vanish as energy increases. Such discrepancies are, however, below our current detection sensibility. As a further example, tritium $\beta$-decay is commonly used for such measurements for its low
endpoint energy and simple nuclear structure (for the case of tritium $\beta$-decay see \cite{23}). Finally, all these aspects have been found in laboratory experiments, where gravitational effects are neglected.

A possible \emph{Gedankenexperiment} is based on a spatial platform on which a baseline is physically constructed  ``near" a maximally-rotating object. The distance between the baseline might be fixed, imposing both $r_A$ and $r_B$, once the distance from the baseline of the compact object is known \emph{a priori}. The idea of a spacial baseline is plausible thanks to the recent developments on the tomography of black holes for example. The previous estimations of angular momentum and mass can give hints on the expected ratios in the correction formulas due to the dipole and quadrupole contributions. In particular the term $\sim J_2^\star\left(1+2M_\star/R_\star\right)\sim 0.03$ if one wants $\Delta \tilde m^2_{21}$ to show a discrepancy of $\sim 0.05$ with respect to $\Delta m^2_{21}$.

As a final discussion about our approach, we can notice that a natural generalization of our results could be argued for non-static space-time. This extension is  possible if one understands how time dependence of the metric would influence neutrino oscillation. For example, in the framework of non-homogeneous space-time such as the Lema\^itre-Tolman-Bondi universe, one involves two generalized functions: the scale-factor 
and the curvature term. This implies that it is necessary to postulate how the functions depend upon $t$ and $r$ otherwise the corresponding integration turns out to be complicated and in many cases impossible to pursue. This limitation is however based on the constraint that, as time goes to infinite, the static results might be recovered as limiting cases. This would fix the boundaries over the extra terms induced within the neutrino oscillation phase. For the above considerations, it is reasonable to assume these corrections would give small deviations from the static case.

\section{Final outlooks and perspectives}\label{concl}

Neutrino oscillations have been investigated in the field of an exterior static axially-symmetric  Zipoy-Voorhees metric, often termed in the literature as: gamma-metric, delta-metric or more frequently $q$-metric. This metric describes a  static and deformed astrophysical object whose gravitational field generalizes the Schwarzschild metric through the inclusion of a quadrupole term.

Particularly, we investigated the consequences of neutrino oscillation on compact object analyzing the weak and strong gravitational regimes, respectively for Solar System, WDs and NSs. For the sake of completeness, we further demonstrated that supernovae alone can not be indicators for deviations from the spherical case of neutrino oscillation. In this case, in particular, neutrinos are produced from the NS born out of the explosion, so neutrino oscillation is also affected by predominant matter effects once neutrinos travel in supernova eject.

Furthermore we showed that
Earth's quadrupole moment negligibly affects the final phase, albeit the quadrupole moment affects the value of $\Delta \bar m_{12}^2$ in the case of rotating WDs and NSs. Thus, specializing our attention to compact objects, we analyzed the dipole and quadrupole cases in view of maximally-rotating configurations and we compared our findings with the ones computed in the Hartle-Thorne and Schwarzschild configurations respectively. Moreover, using experimental data from cosmological probes and nuclear physics experiments, we used the current paradigm purporting the three-flavor neutrino mixing theory. Thence, we computed numerical constraints on survival probability for WDs and NSs, giving basic suggestions toward possible experiments to perform in the next years to check the theoretical deviations here predicted, based on spatial baselines. We showed that for neutrinos detected on Earth the quadrupole moment correction to phase shift is at most $-0.2\%$. From theoretical reasons, we expected this value to be large enough in the field of WDs and NSs. Therefore, we explored this possibility and demonstrated that the angular momentum is crucial for NS, while for the Earth, Sun and WDs the effects of rotation can be neglected with respect to the quadrupolar deformation. In view of the fact that in the following years one can propose the construction of space missions devoted to the direct test of neutrino oscillations in the field of compact objects. The results obtained here will be studied in a more general and complicated space-time.

\section*{Acknowledgments} 

K.B. acknowledges the support provided by the Ministry of Education and Science of the Republic of Kazakhstan Target Program ’Center of Excellence for Fundamental and Applied Physics’ IRN: BR05236454, Grant IRN: AP08052311, Grant IRN: AP05134454. M.M. acknowledges the support of INFN, Frascati National Laboratories, for Iniziative Specifiche MOONLIGHT2.

\end{document}